\def\~{{$\tilde{\phantom{a}}$}}

\documentclass [12pt] {article}
\usepackage{epsfig}
\usepackage{color}

\def\thebibliography#1{\section{References}\markboth
 {REFERENCES}{REFERENCES}\list
 {[\arabic{enumi}]}{\settowidth\labelwidth{[#1]}\leftmargin\labelwidth
 \advance\leftmargin\labelsep
 \usecounter{enumi}}
 \def\newblock{\hskip .11em plus .33em minus -.07em}
 \sloppy
 \sfcode`\.=1000\relax}
\def\upcite#1{\raise6pt\hbox{\scriptsize
\cite{#1}}}
  \def\lsim{\mathrel {\vcenter {\baselineskip 0pt \kern 0pt
    \hbox{$<$} \kern 0pt \hbox{$\sim$} }}}
    \def\gsim{\mathrel {\vcenter {\baselineskip 0pt \kern 0pt
    \hbox{$>$} \kern 0pt \hbox{$\sim$} }}}

\def\hline{\noalign{\hrule \vskip2pt}}

\def\|{\ifmmode\Vert\else \char`\|\fi}
\ifx\oldzeta\undefined                          
  \let\oldzeta=\zeta                            
  \def\zzeta{{\raise 2pt\hbox{$\oldzeta$}}}     
  \let\zeta=\zzeta                              
\fi

\ifx\oldchi\undefined                           
  \let\oldchi=\chi                              
  \def\cchi{{\raise 2pt\hbox{$\oldchi$}}}       
  \let\chi=\cchi                                
\fi

\def\frac#1#2{{#1 \over #2}}

\def\half{\ifinner {\scriptstyle {1 \over 2}}
   \else {1 \over 2} \fi}

\def\simge{\mathrel{%
   \rlap{\raise 0.511ex \hbox{$>$}}{\lower 0.511ex \hbox{$\sim$}}}}
\def\simle{\mathrel{
   \rlap{\raise 0.511ex \hbox{$<$}}{\lower 0.511ex \hbox{$\sim$}}}}

\def\buildchar#1#2#3{{\null\!                   
   \mathop#1\limits^{#2}_{#3}                   
   \!\null}}                                    
\def\overcirc#1{\buildchar{#1}{\circ}{}}

\def\slashchar#1{\setbox0=\hbox{$#1$}           
   \dimen0=\wd0                                 
   \setbox1=\hbox{/} \dimen1=\wd1               
   \ifdim\dimen0>\dimen1                        
      \rlap{\hbox to \dimen0{\hfil/\hfil}}      
      #1                                        
   \else                                        
      \rlap{\hbox to \dimen1{\hfil$#1$\hfil}}   
      /                                         
   \fi}                                         %

\def\subrightarrow#1{
  \setbox0=\hbox{
    $\displaystyle\mathop{}
    \limits_{#1}$}
  \dimen0=\wd0
  \advance \dimen0 by .5em
  \mathrel{
    \mathop{\hbox to \dimen0{\rightarrowfill}}
       \limits_{#1}}}                           

\def\overlay#1#2{\ifmmode%
\setbox0=\hbox{$#1$}%
\setbox1=\hbox to\wd0{\hss$#2$\hss}\else%
\setbox0=\hbox{#1}%
\setbox1=\hbox to\wd0{\hss#2\hss}\fi%
#1\hskip-\wd0\box1 }

\def\pmb#1{\leavevmode\setbox0=\hbox{#1}%
\kern-.02em\copy0\kern-\wd0
\kern.04em\copy0\kern-\wd0
\kern-.02em\raise.04em\box0 }

\def\vereq#1#2{\lower3pt\vbox{\baselineskip1.5pt \lineskip1.5pt
\ialign{$\m@th#1\hfill##\hfil$\crcr#2\crcr\sim\crcr}}}

\def\tensor#1{\protect\@ontopof{#1}{\leftrightarrow}{1.15}\mathord{\box2}}
\def\overstar#1{\protect\@ontopof{#1}{\ast}{1.15}\mathord{\box2}}
\def\overdots#1{\protect\@ontopof{#1}{\cdots}{1.0}\mathord{\box2}}
\def\overcirc#1{\protect\@ontopof{#1}{\circ}{1.2}\mathord{\box2}}
\def\loarrow#1{\protect\@ontopof{#1}{\leftarrow}{1.15}\mathord{\box2}}
\def\roarrow#1{\protect\@ontopof{#1}{\rightarrow}{1.15}\mathord{\box2}}

\def\@ontopof#1#2#3{%
{\mathchoice
{\@@ontopof{#1}{#2}{#3}\displaystyle\scriptstyle}%
{\@@ontopof{#1}{#2}{#3}\textstyle\scriptstyle}%
{\@@ontopof{#1}{#2}{#3}\scriptstyle\scriptscriptstyle}%
{\@@ontopof{#1}{#2}{#3}\scriptscriptstyle\scriptscriptstyle}%
}%
}

\def\@@ontopof#1#2#3#4#5{%
\setbox0=\hbox{$#4#1$}%
\setbox1=\hbox{$#5#2$}%
\setbox2=\hbox{}\ht2=\ht0 \dp2=\dp0 %
\ifdim\wd0>\wd1 %
\setbox1=\hbox to\wd0{\hss\box1\hss}%
\mathord{\rlap{\raise#3\ht0\box1}\box0}%
\else   %
\setbox1=\hbox to.9\wd1{\hss\box1\hss}%
\setbox0=\hbox to\wd1{\hss$#4\relax#1$\hss}%
\mathord{\rlap{\copy0}\raise#3\ht0\box1}%
\fi
}%

\def\lambdabar{\protect\@lambdabar}
\def\@lambdabar{%
\relax
\bgroup
\def\@tempa{\hbox{\raise.73\ht0
\hbox to0pt{\kern.25\wd0\vrule width.5\wd0
height.1pt depth.1pt\hss}\box0}}%
\mathchoice{\setbox0\hbox{$\displaystyle\lambda$}\@tempa}%
{\setbox0\hbox{$\textstyle\lambda$}\@tempa}%
{\setbox0\hbox{$\scriptstyle\lambda$}\@tempa}%
{\setbox0\hbox{$\scriptscriptstyle\lambda$}\@tempa}%
\egroup
}

\def\corresponds{{\lower.2ex\hbox{=}}{\rm\kern-.75em^\triangle}}
\def\succsim{\succ\kern-.9em_\sim\kern.3em}
\def\precsim{\prec\kern-1em_\sim\kern.3em}
\def\slantfrac#1#2{\kern1em^{#1}\kern-.3em/\kern-.1em_{#2}}

\begin{document}
                                                                
\begin{center}
{\Large\bf Time-Reversed Diffraction}
\\

\medskip

Max. S.~Zolotorev
\\
{\sl Center for Beam Physics, Lawrence Berkeley National Laboratory,
Berkeley, CA 94720}
\\
Kirk T.~McDonald
\\
{\sl Joseph Henry Laboratories, Princeton University, Princeton, NJ 08544}
\\
(Sep.\ 5, 1999)
\end{center}

\section{Problem}

In the usual formulation of the Kirchhoff diffraction integral, a scalar field 
with harmonic time dependence at frequency $\omega$ is deduced at the interior 
of a charge-free volume from knowledge of the field (or its normal derivative) 
on the bounding surface.  In particular, the field is propagated forwards in 
time from the boundary to the desired observation point.

Construct a time-reversed version of the Kirchhoff integral in which the 
knowledge of the field on the boundary is propagated backwards in time into 
the interior of the volume.

Consider the example of an optical focus at the origin for a system with the 
$z$ axis as the optic axis.
In the far field beyond the focus a Gaussian beam has cone angle 
$\theta_0 \equiv \sqrt{2} \sigma_\theta$, and the $x$ component
of the electric field in a spherical coordinate system is given approximately by
\begin{equation}
E_x(r,\theta,\phi,t) = E(r) e^{i(kr - \omega t)} e^{-\theta^2 / \theta_0^2},
\label{eq1}
\end{equation}
where $k = \omega/c$ and $c$ is the speed of light.  
Deduce the field near the focus.

Since the Kirchhoff diffraction formalism requires the volume to be charge 
free, the time-reversed technique is
not applicable to cases where the source of the field is inside the volume.   
Nonetheless, the reader may find it instructive to attempt to apply the 
time-reversed diffraction integral to the example of an oscillating dipole at 
the origin.

\section{The Kirchhoff Integral via Green's Theorem}

A standard formulation of Kirchhoff's diffraction integral for a scalar field 
$\psi({\bf x})$ with time dependence $e^{-i \omega t}$ is
\begin{equation}
\psi({\bf x}) \approx {k \over 2 \pi i} \int_S {e^{i k r'} \over r'} 
\psi({\bf x}') d{\rm Area}',
\label{eql19}
\end{equation}
where the spherical waves $e^{i(k r' - \omega t)} / r'$ are outgoing, and
$r'$ is the magnitude of vector ${\bf r}' = {\bf x} - {\bf x}'$.

For a time-reversed formulation in which we retain the time dependence as 
$e^{-i \omega t}$, the spherical waves of interest are the incoming waves 
$e^{-i(kr' + \omega t)} / r'$.  In brief, the desired time-reversed diffraction 
integral is obtained from eq.~(\ref{eql19}) on replacing $i$ by $-i$:
\begin{equation}
\psi({\bf x}) \approx {ik \over 2 \pi} \int_S {e^{-i k r'} \over r'} 
\psi({\bf x}') d{\rm Area}'.
\label{eql20}
\end{equation}

For completeness, we review the derivation of eqs.~(\ref{eql19})-(\ref{eql20})
via Green's theorem.  See also, sec.~10.5 of ref.~\cite{Jackson}.

Green tells us that for any two well-behaved scalar fields $\phi$ and $\psi$,
\begin{equation}
\int_V (\phi \nabla^2 \psi - \psi \nabla^2 \phi) d{\rm Vol} =
\int_S (\phi \nabla' \psi - \psi \nabla' \phi) \cdot d{\bf S}'.
\label{eql9}
\end{equation} 
The surface element $d{\bf S}'$ is directly outward from surface $S$.
We consider fields with harmonic time dependence at frequency $\omega$, and 
assume the factor $e^{-i \omega t}$.  The wave function of interest, $\psi$, 
is assumed to have no
sources within volume $V$, and so obeys the Helmholtz wave equation,
\begin{equation}
\nabla^2 \psi + k^2 \psi = 0.
\label{eql10}
\end{equation}

We choose function $\phi({\bf x})$ to correspond to waves associated with a 
point source at ${\bf x}'$.  That is,
\begin{equation}
\nabla^2 \phi + k^2 \phi = - \delta^3({\bf x} - {\bf x}').
\label{eq11}
\end{equation}
The well-known solutions to this are the incoming and outgoing spherical waves,
\begin{equation}
\phi_\pm({\bf x},{\bf x}') = {e^{\pm i k r'} \over r'},
\label{eql12}
\end{equation}
where the 
+ sign corresponds to the outgoing wave.
We recall that
\begin{equation}
\nabla' r' = -{{\bf r}' \over r'} = - \hat {\bf n}_o,
\label{eql13}
\end{equation}
where $\hat{\bf n}_o$ points towards the observer at {\bf x}.  Then,
\begin{equation}
\nabla' \phi_\pm = \mp i k \hat{\bf n}_o \left( 1 \pm {1 \over i k r'} \right) 
\phi.
\label{eql14}
\end{equation}
Inserting eqs.~(\ref{eql10})-(\ref{eql14}) into eq.~(\ref{eql9}), we find
\begin{equation}
\psi({\bf x}) = - {1 \over 4 \pi} \int_S {e^{\pm i k r'} \over r'} 
\hat{\bf n}' \cdot \left[ \nabla' \psi \pm i k \hat{\bf n}_o 
\left( 1 \pm {1 \over i k r'} \right) \psi \right] d{\rm Area}',
\label{eql15}
\end{equation}
where the overall minus sign holds with the convention that $\hat{\bf n}'$ is 
the inward normal to the surface.

We only consider cases where the source of the wave $\psi$ is far from the 
boundary surface, so that on the boundary $\psi$ is well approximated as a 
spherical wave,
\begin{equation}
\psi({\bf x}') \approx A {e^{i k r_s} \over r_s},
\label{eql16}
\end{equation}
where $r_s$ is the magnitude of the vector ${\bf r}_s = {\bf x}' - {\bf x}_s$ 
from the effective source point ${\bf x}_s$ to the point ${\bf x}'$ on the 
boundary surface.  In this case,
\begin{equation}
\nabla' \psi = i k \hat{\bf n}_s \left( 1 \pm {1 \over i k r_s} \right) \psi,
\label{eql17}
\end{equation}
where $\hat{\bf n}_s = {\bf r_s}/r_s$

We also suppose that the observation point is far from the boundary surface,
so that $k r' \ll 1$ as well as $k r_s \ll 1$.  Hence, we neglect the terms in 
$1/i k r'$ and $1 /i k r_s$ to find
\begin{equation}
\psi({\bf x}) = - {ik \over 4 \pi} \int_S {e^{\pm i k r'} \over r'} \hat{\bf n}'
\cdot ( \hat{\bf n}_s \pm \hat{\bf n}_o)  \psi({\bf x}') d{\rm Area}'.
\label{eql18}
\end{equation}

The usual formulation, eq.~(\ref{eql19}), of Kirchhoff's law is obtained using 
outgoing waves (+ sign), and
the paraxial approximation that $\hat {\bf n}' \approx \hat {\bf n}_o \approx
\hat {\bf n}_s$.  The latter tacitly assumes that the effective source is 
outside volume $V$. 

Here, we are interested in the case where the effective source is inside the 
volume $V$, so that the paraxial approximation is $\hat {\bf n}' \approx 
\hat {\bf n}_o \approx - \hat {\bf n}_s$.  When we use the incoming wave 
function to reconstruct $\psi({\bf x},t)$ from information on the boundary at 
time $t' > t$, we use the $-$ sign in eq.~(\ref{eql18})
to find eq.~(\ref{eql20}).

Note that in this derivation, we assumed that $\psi$ obeyed eq.~(\ref{eql10}) 
throughout volume $V$, and so the actual source of $\psi$ cannot be within $V$.
Our time-reversed Kirchhoff integral (\ref{eql20}) can only be applied when any
source inside $V$ is virtual.  This includes the interesting case of a focus 
of an optical system (secs.~4 and 5).  
However, we cannot expect  eq.~(\ref{eql20}) to 
apply to the case of a physical source, such as an oscillating dipole, inside 
volume $V$ (sec.~6).   The laws of diffraction do not permit electromagnetic 
waves to converge into a volume smaller than a wavelength cubed, and so
eq.~(\ref{eql20}) cannot be expected to describe the near fields around a 
source smaller than this.

\section{A Plane Wave}

The time-reversed Kirchhoff integral (\ref{eql20}) for the $x$ component of 
the electric field is
\begin{equation}
E_x({\rm obs,now}) = {i k \over 2 \pi} \int {e^{-ikr'} \over r'}
 E_x(r,\theta,\phi, \rm future) d{\rm Area} ,
\label{eql1}
\end{equation} 
where $r'$ is the distance from the observation point
${\bf r}_{\rm obs} = (x,y,z)$ in rectangular coordinates to a point
${\bf r} = r (\sin\theta \cos\phi, \sin\theta \sin\phi, \cos\theta)$ on a 
sphere of radius $r$ in the far field.

As a first example, consider a plane electromagnetic wave,
\begin{equation}
E_x = E_0 e^{i(kz - \omega t)} = E_0 e^{i(kr \cos\theta - \omega t)},
\label{3.1}
\end{equation}
where the second form holds in a spherical coordinate system $(r,\theta,\phi)$
where $\theta$ is measured with respect to the $z$ axis.
We take the point of observation to be $(x,y,z) = (0,0,r_0)$, and evaluate
the diffraction integral (\ref{eql1}) over a sphere of radius $r \gg r_0$.
In the exponential factor in the Kirchhoff integral, we approximate $r'$ as
\begin{equation}
r' \approx r - \hat{\bf r} \cdot {\bf r}_{\rm obs}
= r - r_0 \cos\theta,
\label{eq3.2}
\end{equation}
while in the denominator we approximate $r'$ as $r$.  Then,
\begin{eqnarray}
E_x({\rm obs}) & \approx & {ik \over 2 \pi} \int_{-1}^1 r^2 d\cos\theta
\int_0^{2 \pi} d \phi \ {e^{-ik(r - r_0 \cos\theta)} \over r} 
E_0 e^{ikr \cos\theta}
\nonumber \\
& = & {r \over r + r_0} E_0 [e^{ikr_0} - e^{-ik(2 r + r_0)} ]
\label{3.3} \\
& \approx & E_0 e^{i k r_0},
\nonumber
\end{eqnarray}
where we ignore the rapidly oscillating term $e^{-ik(2 r + r_0)}$ as unphysical.

This verifies that the time-reversed diffraction formula works for a simple
example.

\section{The Transverse Field near a Laser Focus}

We now consider the far field of a laser beam whose optic axis is the $z$ axis
with focal point at the origin.  The polarization is along the $x$ axis, and the
electric field has
Gaussian dependence on polar angle with characteristic angle $\theta_0 \ll 1$.
Then, we can write
\begin{equation}
E_x(r,\theta,\phi) = E(r) e^{ikr} e^{-\theta^2 / \theta_0^2},
\label{eql2}
\end{equation}
where $E(r)$ is the magnitude of the electric field on the optic axis at 
distance $r$ from the focus.
In the exponential factor in the Kirchhoff integral (\ref{eql1}), 
 $r'$ is the distance from the observation
point ${\bf r}_{|rm obs} = (x,y,z)$ to a point
${\bf r } = r(\sin\theta \cos\phi, \sin\theta \sin\phi, \cos\theta)$ on the 
sphere.  We approximate $r'$ as
\begin{equation}
r' \approx r - \hat{\bf r} \cdot {\bf r}_{\rm obs}
= r - x \sin\theta \cos\phi - y \sin\theta \sin\phi - z \cos\theta ,
\label{eql3}
\end{equation}
while in the denominator we approximate $r'$ as $r$.
Inserting eqs.~(\ref{eql2}) and (\ref{eql3}) into (\ref{eql1}), we find
\begin{eqnarray}
E_x({\rm obs}) 
& = & {i k r E(r) \over 2 \pi} 
\int_{-1}^1 e^{i k z \cos\theta} e^{-\theta^2 / \theta_0^2} d\cos\theta 
\int_0^{2\pi} e^{i k x \sin\theta \cos\phi + i k y \sin\theta \sin\phi} d\phi
\nonumber \\
& = & i k r E(r) \int_{-1}^1 e^{i k z \cos\theta} e^{-\theta^2 /  \theta_0^2}
J_0(k \rho \sin\theta) d\cos\theta,
\label{eql4}
\end{eqnarray}
where 
\begin{equation}
\rho = \sqrt{x^2 + y^2},
\label{eq4a}
\end{equation}
and $J_0$ is the Bessel function of order zero.

\begin{figure}[htp]  
\begin{center}
\includegraphics[width=4in, angle=0, clip]{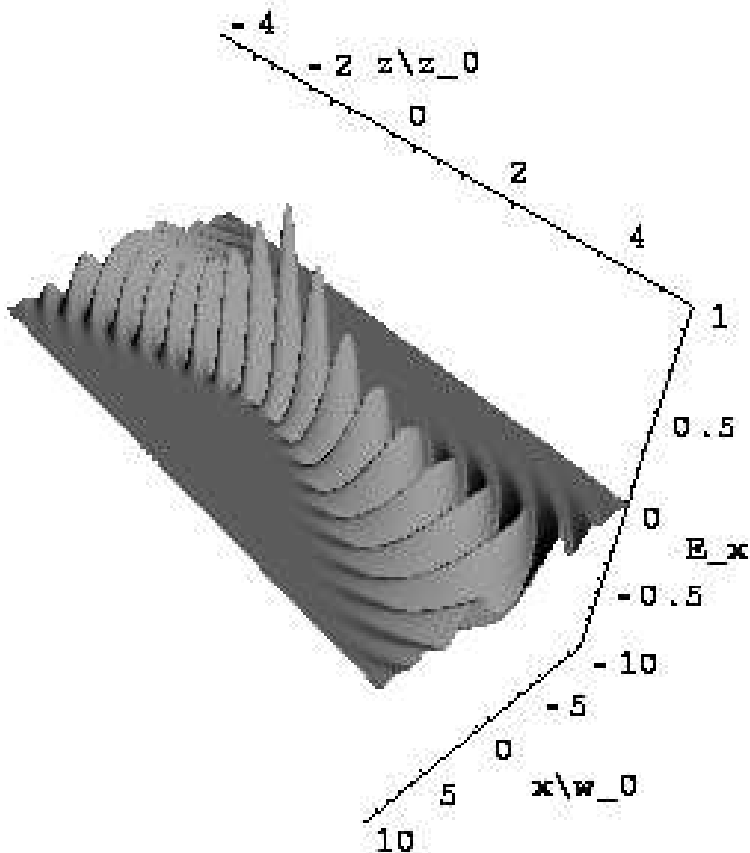}
\parbox{5.5in} 
{\caption[ Short caption for table of contents ]
{\label{exl} The electric field $E_x(x,0,z)$ of a linearly polarized Gaussian
beam with diffraction angle $\theta_0 = 0.45$.
}}
\end{center}
\end{figure}

Since we assume that the characteristic angle $\theta_0$ of the laser beam is 
small, we can approximate $\cos\theta$ as $1 - \theta^2/2$ and 
$k \rho \sin\theta$ as $k \rho \theta$.  
Then, 
we have
\begin{eqnarray}
E_x({\rm obs}) 
& \approx & i k r E(r) e^{i k z} 
\int_0^\infty e^{-(2 / \theta_0^2 + i k z) \theta^2 / 2}
J_0\left( \sqrt{2} k \rho \sqrt{\theta^2/2} \right) d(\theta^2/2) 
\nonumber \\
& = &
{i k \theta_0^2 r E(r) e^{i k z} 
e^{- k^2 \theta_0^2 \rho^2 / 4 (1 + i k \theta_0^2 z /2)}
\over 2 (1 + i k \theta_0^2 z /2)},
\label{eql5}
\end{eqnarray}
where the Laplace transform, which is given explicitly in \cite{Magnus},
 can be evaluated using 
the series expansion for the Bessel function.
This expression can be put in a more familiar form by introducing the
Rayleigh range (depth of focus),
\begin{equation}
z_0 = {2 \over k \theta_0^2},
\label{eql6}
\end{equation}
and the so-called waist of the laser beam,
\begin{equation}
w_0 = \theta_0 z_0 = {2 \over k \theta_0}.
\label{eql6a}
\end{equation}
We define the electric field strength at the focus $(\rho = 0,z = 0)$
to be $E_0$, so we learn that the far-field strength is related by
\begin{equation}
E(r) = -i {z_0 \over r} E_0.
\label{eql7}
\end{equation}
The factor $-i = e^{-i \pi / 2}$ is the $90^\circ$ Guoy phase shift between 
the focus and the far field.  Then, the transverse component of the
electric field near the focus is
\begin{eqnarray}
E_x(x,y,z) & \approx & E_0 {e^{-\rho^2 / w_0^2 (1 + i z / z_0)} e^{i k z} 
\over (1  + i z / z_0)}
\nonumber \\
& = & E_0 {e^{-\rho^2 / w_0^2 (1 + z^2 / z_0^2)}
e^{-i\tan^{-1}z/z_0} e^{i\rho^2 z / w_0^2 z_0 (1 + z^2 / z_0^2)}
 e^{i k z} \over \sqrt{1 + (z/z_0)^2}}.
\label{eql8}
\end{eqnarray}
This is the usual form for the lowest-order mode of a linearly polarized
 Gaussian laser beam \cite{Eberly}.  Figure \ref{exl} plots this field.

The Gaussian beam (\ref{eql8}) could also be deduced by a similar argument
using eq.~(\ref{eql19}), starting from the far field of the laser before the
focus.  The form (\ref{eql8}) is symmetric in $z$ except for a phase factor,
and so is a solution to the problem of transporting a wave from $z = -r$ to
$z = +r$ such that the functional dependence on $\rho$ and $z$ is invariant up 
to a phase factor.  One of the earliest derivations \cite{Boyd}
 of the Gaussian beam 
was based on the formulation of this problem as an integral equation 
for the eigenfunction (\ref{eql8}).

\section{The Longitudinal Field}


Far from the focus, the electric field {\bf E}({\bf r}) is perpendicular to the 
radius vector {\bf r}.  For a field linearly polarized in the $x$ direction,
there must also be a longitudinal component $E_z$ related by
\begin{equation}
{\bf E} \cdot \hat {\bf r} = E_x \sin\theta \cos\phi + E_z \cos\theta = 0.
\label{eql201}
\end{equation}
Thus, far from the focus,
\begin{equation}
E_z({\bf r}) = - E_x({\bf r}) \tan\theta \cos\phi.
\label{eql202}
\end{equation}
Then, similarly to eqs.~(\ref{eql1}) and (\ref{eql4}), we have
\begin{eqnarray}
E_z({\rm obs})
& = & {i k \over 2 \pi} \int {e^{-ikr'} \over r'}  E_z({\bf r}) d{\rm Area} 
\nonumber \\
& = & -{i k r E(r) \over 2 \pi} 
\int_{-1}^1 
e^{i k z \cos\theta} e^{-\theta^2 / \theta_0^2} \tan\theta 
d\cos\theta 
\int_0^{2\pi} e^{i k x \sin\theta \cos\phi + i k y \sin\theta \sin\phi} 
\cos\phi d\phi
\nonumber \\
& = & - {i k x z_0 E_0 \over \rho} \int_{-1}^1 
e^{i k z \cos\theta} e^{-\theta^2 /  \theta_0^2} \tan\theta 
J_1(k \rho \sin\theta) d\cos\theta,
\label{eql203}
\end{eqnarray}
using eq.~(3.937.2) of \cite{Gradshteyn}.

\begin{figure}[htp]  
\begin{center}
\includegraphics[width=4in, angle=0, clip]{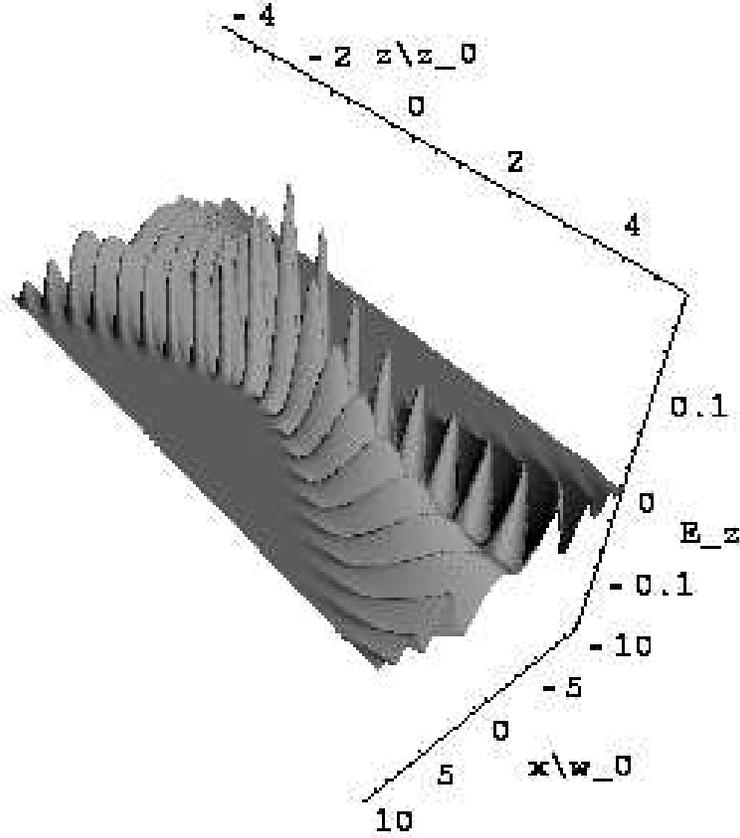}
\parbox{5.5in} 
{\caption[ Short caption for table of contents ]
{\label{ezl} The electric field $E_z(x,0,z)$ of a linearly polarized Gaussian
beam with diffraction angle $\theta_0 = 0.45$.
}}
\end{center}
\end{figure}

We again note that the integrand is significant only for small $\theta$, so
we can approximate eq.~(\ref{eql203}) as the Laplace transform
\begin{eqnarray}
E_z(x,y,z) 
& \approx & -i k^2 x z_0 E_0 e^{i k z} 
 \sqrt{2} \int_0^\infty e^{-(2 / \theta_0^2 + i k z) \theta^2 / 2}
\sqrt{\theta^2/2} J_1\left( \sqrt{2}k \rho \sqrt{\theta^2/2} \right) 
d(\theta^2/2) 
\nonumber \\
& = &
-{i k^2 \theta_0^4 x z_0 E_0 e^{i k z} e^{- \rho^2 / w_0^2 (1 + i z /z_0)}
\over 4 (1 + i z /z_0)^2}
\nonumber \\
& = &
-i \theta_0 {x \over w_0} {E_x(x,y,z) \over (1 + i z /z_0)},
\label{eql204}
\end{eqnarray}
with $E_x$ given by eq.~(\ref{eql8}).   Figure \ref{ezl} plots this field.

Together, the electric field components given by eqs.~(\ref{eql8}) and
(\ref{eql204}) satisfy the Maxwell equation $\nabla \cdot {\bf E} = 0$ to
order $\theta_0^2$ \cite{Lax,Davis,Barton}.

\section{Oscillating Dipole at the Origin}

We cannot expect the Kirchhoff diffraction integral to apply to the example of 
an oscillating dipole, if our bounding surface surrounds the dipole.
Let us see what happens if we try to use eq.~(\ref{eql20}) anyway.

The dipole is taken to be at the origin, with  moment $p$ along the $x$ axis. 
Then, the $x$ component of the radiation field is
\begin{equation}
E_x = k^2 p \sin\theta_x {e^{ikr} \over r}.
\label{eqk1}
\end{equation}
where $\theta_x$ is the angle between the $x$ axis and a radius vector to the
observer.
We consider an observer near the origin at $(x,y,z) = (0,0,r_0)$,  for which
$\sin\theta_x = 1$, and so
\begin{equation}
E_x({\rm obs}) = k^2 p {e^{ikr_0} \over r_0}.
\label{eqk2}
\end{equation}

We now attempt to reconstruct this field near the origin from its value on a 
sphere of radius $r$ using the time-reversed Kirchhoff integral (\ref{eql20}).
We use a spherical coordinate system $(r,\theta,\phi)$ that favors the $z$ axis.
Then, the $x$ component of the radiation field on the sphere of radius $r$ is
\begin{equation}
E_x(r,\theta,\phi) = k^2 p \sqrt{1 - \sin^2\theta \cos^2\phi}  
{e^{ikr} \over r}.
\label{eqk4}
\end{equation}
This form cannot be integrated analytically, so we use a Taylor expansion 
of the square root, which will lead to an expansion in powers of $1/r_0$.  
It turns out that the coefficient of the $1/r_0$ term, which is our main 
interest, is very close to that if we simply approximate the square root by 
unity.  For brevity, we write
\begin{equation}
E_x(r,\theta,\phi) \approx k^2 p {e^{ikr} \over r}.
\label{eqk4a}
\end{equation}

In the time-reversed Kirchhoff integral (\ref{eql20}),
we make the usual approximation that $r' = r - r_0 \cos\theta$ in the
exponential factor, but $r' = r$ in the denominator.  Then, using 
eq.~(\ref{eqk4a}) we have
\begin{eqnarray}
E_x({\rm obs}) & \approx & {i k^3 p e^{-ikr} \over 2 \pi r} 
\int_{-1}^1 r^2 d\cos\theta 
\int_0^{2\pi}  d\phi e^{ikr_0 \cos\theta}  {e^{ikr} \over r}
\nonumber \\  
& = & k^2 p {e^{i k r_0} \over r_0} - k^2 p {e^{-i k r_0} \over r_0} 
\nonumber \\  
& = & 2 i k^3 p {\sin k r_0 \over k r_0}.
\label{eqk6}
\end{eqnarray}
The first, outgoing wave in middle line of eq.~(\ref{eqk6}) is the desired 
form, but the second, incoming wave is of the same magnitude.  Together, they 
lead to the form $\sin(kr_0)/k r_0$ which is nearly constant for 
$k r_0 \lsim 1$.  The presence of outgoing as well as incoming waves is to be
expected because dipole radiation is azimuthally symmetric about the $x$ axis.
In the absence of a
charged source at the origin, an outgoing wave at $\theta = \pi$ must 
correspond to an incoming wave at $\theta = 0$.

The result that the reconstructed field is uniform for distances within a 
wavelength of the origin is consistent with the laws of diffraction that 
electromagnetic waves cannot be focused to a region smaller than a wavelength.
Far fields of the form (\ref{eqk1}) could only be propagated back to the form 
of dipole fields near the origin with the addition of nonradiation fields tied 
to a charge at the origin.  Such a construction is outside the scope of optics 
and diffraction.



\end{document}